\documentstyle[12pt,scipp,epsfig]{article}
\begin{document}

\renewcommand{\thefootnote}{\fnsymbol{footnote}}

\def\gtwid{\mathrel{\raise.3ex\hbox{$>$\kern-.75em\lower1ex\hbox{$\sim$}}}}
\def\ltwid{\mathrel{\raise.3ex\hbox{$<$\kern-.75em\lower1ex\hbox{$\sim$}}}}
\def\gsim{\gtwid}
\def\lsim{\ltwid}
\def \hMpc{\, h^{-1}\, \rm Mpc}
\def \kms{\, {\rm km \, s}^{-1}}
\def \kmsMpc{\, {\rm km \, s}^{-1} \, {\rm Mpc}}
\def\ie{{\it i.e.}}
\widowpenalty=100000
\begin{center}
{\Large 
Status of Cosmological Parameters:
$\Omega_0\approx 0.3$ vs. $\Omega=1$
\footnote{To appear in {\it International School of Physics ``Enrico
Fermi''}, Course CXXXII: Dark Matter in the Universe, Varenna 1995, eds.
S. Bonometto, J.R. Primack, A. Provenzale.}
}\\[2pc]

{Joel R. Primack} \\[3pt]
{Santa Cruz Institute for Particle Physics,\\
 University of California, Santa Cruz, CA~95064}\\[3pc]

 Abstract \\[1pc]
\begin{quote}
\small
The cosmological parameters that I will discuss are the
traditional ones: the Hubble parameter $H_0 \equiv 100 h$
km s$^{-1}$ Mpc$^{-1}$, the age of the universe $t_0$, the
average density $\Omega_0 \equiv \bar\rho/\rho_c$ in units
of critical density $\rho_c$, and the cosmological constant
$\Lambda$ (or $\Omega_\Lambda \equiv \Lambda/(3H_0^2)$).
To focus the discussion, I will concentrate on
the issue of the value of $\Omega_0$ in
currently popular cosmological models in which most of the
dark matter is cold, especially Cold + Hot Dark Matter
(CHDM), and flat ($\Omega_0 + \Omega_\Lambda=1$)
low-$\Omega$ CDM with a Cosmological
Constant ($\Lambda$CDM).  The evidence would favor a small
$\Omega_0 \approx 0.3$ if (1) the Hubble parameter actually
has the high value $h \approx 0.75$ favored by many
observers, and the age of the universe $t_0 \geq 13$ Gy; or
(2) the baryonic/total mass ratio in clusters of galaxies
is actually $\sim
15$\%, about 3 times larger than expected for standard Big
Bang Nucleosynthesis in an $\Omega=1$ universe, and standard
BBN is actually right in predicting that the density of ordinary
matter is $\Omega_b \approx 0.0125 h^{-2}$, based mainly on $^4$He 
and $7$Li data, despite the recent measurement by Tytler of 
$D/H=2.4\times 10^{-5}$ in two high-redshift Lyman limit 
systems, implying $\Omega_b\approx 0.024 h^{-2}$.  
The evidence would favor $\Omega=1$ if (1) the POTENT 
analysis of galaxy peculiar velocity data is right, in 
particular regarding outflows from voids or the inability to 
obtain the present-epoch non-Gaussian density distribution 
from Gaussian initial fluctuations in a low-$\Omega$ 
universe; or (2) the preliminary report from LSND indicating 
a neutrino mass $\gsim 2.4$ eV is right, since that would be 
too much hot dark matter to allow significant structure 
formation in a low-$\Omega_0$ $\Lambda$CDM model.  
Statistics on gravitational lensing of quasars provide a 
strong upper limit on $\Lambda$. It also appears to be 
possible to measure the deceleration parameter 
$q_0=\Omega_0/2-\Omega_\Lambda$ on very large scales using 
high-redshift Type Ia supernovae; the preliminary results 
suggest that $\Omega_0 \sim 1$ and $\Omega_\Lambda$ is 
small.  The era of structure formation is another important 
discriminant between these alternatives, low $\Omega_0$ 
favoring earlier structure formation, and $\Omega=1$ 
favoring later formation with many clusters and larger-scale 
structures still forming today. Reliable data on all of 
these issues is becoming available so rapidly today that 
there is reason to hope that a clear decision between these 
alternatives will be possible within the next few years. 
\end{quote} \end{center} 

\vfill\clearpage

\section{Introduction}

As I write this in spring 1996 \cite{Snowmass95}, there is
still concern about a crisis in cosmology.  The first
article \cite{Freedman94} using HST observations of Cepheid
variable stars to determine a distance to a relatively
distant galaxy, the beautiful face-on spiral M100, was
published about a year ago.  The distance obtained was $17.1
\pm 1.8$ Mpc.  With the additional assumptions that M100
lies in the core of the Virgo cluster and that the recession
velocity of Virgo corrected for infall is about 1400 km
s$^{-1}$, the value obtained for the Hubble parameter is at
the high end of recent estimates: $H_0=80 \pm 17 \kmsMpc$.
Using $h=0.8$ gives, for $\Omega=1$ and a vanishing
cosmological constant $\Lambda=0$, a very short age for the
universe $t_0=8.15$ Gyr, almost certainly younger than the
ages of Milky Way globular clusters and even some nearby
white dwarfs.  Even with $\Omega_0=0.3$, about as low as
permitted by observations, and with $\Omega_\Lambda \equiv
\Lambda/(3H_0^2) = 0.7$, perhaps even higher than current
data allow, $t_0= 11.8$ Gyr for $h=0.8$, which is also
uncomfortably short.  Is this a crisis?  Does it undermine
the strong evidence for the standard Big Bang?  I don't
think so.  Given the considerable uncertainties reflected in
the large quoted error on $H_0$, I think even $\Omega=1$
models are not excluded.  But this Cepheid measurement of
the distance to M100 bodes well for the success of the HST
Key Project on the Extragalactic Distance Scale, which seeks
to measure $H_0$ to 10\% within a few years.  There has
also been recent progress in using Type Ia supernovae as
distance indicators, for measuring both $H_0$ and the
deceleration parameter $q_0=\Omega_0/2-\Omega_\Lambda$.  The
expectation that accurate measurements of the key
cosmological parameters will soon be available is great news
for theorists trying to construct a fundamental theory of
cosmology, and helps motivate the present summary.

In addition to the Hubble parameter $H_0 \equiv 100 h$
km s$^{-1}$ Mpc$^{-1}$, I will discuss the age of the
universe $t_0$, the average density $\Omega_0$, and
the cosmological constant $\Lambda$.  But there are several
additional cosmological parameters whose values are critical
for modern theories: the densities of ordinary matter
$\Omega_b$, cold dark matter $\Omega_c$, and hot dark matter
$\Omega_\nu$, and, for primordial fluctuation spectra
$P(k)=Ak^{n_p}$, the index $n_p$ and the amplitude $A$, or
equivalently (for a given model) the bias parameter $b
\equiv 1/\sigma_8$, where $\sigma_8 \equiv (\delta
M/M)_{rms}$ on a scale of $8 \hMpc$.  A full treatment of
these parameters would take a much longer article than this
one, so to focus the discussion I will concentrate on the
issue of the value of the density $\Omega_0$ in currently
popular cosmological models in which most of the dark matter
is cold.  Although much of the following discussion will be
quite general, it will be helpful to focus on two specific
cosmological models which are perhaps the most popular today
of the potentially realistic models: low-$\Omega$ Cold Dark
Matter with a Cosmological Constant ($\Lambda$CDM, discussed
as an alternative to $\Omega=1$ CDM since the beginning of
CDM \cite{BFPR84,Peeb84}, and worked out in detail in
\cite{CenOstGn}), and $\Omega=1$ Cold + Hot Dark Matter
(CHDM, proposed in 1984 \cite{CHDM84}, and first worked out
in detail in 1992-3 \cite{DSS92,KHPR}).  I will begin by
summarizing the rationale for these models.

\section{Models with Mostly Cold Dark  Matter}

Let me begin here by recalling the definitions of ``hot''
and ``cold'' dark matter.  These terms describe the
astrophysically relevant aspects of candidate dark matter
particles.  The fact that the observational lower bound on
$\Omega_0$ --- namely $0.3 \lsim \Omega_0$ --- exceeds the
most conservative upper
limit on baryonic mass $\Omega_b \lsim 0.03 h^{-2}$ from Big
Bang Nucleosynthesis \cite{Copi95} is the main evidence that
there must be such nonbaryonic dark matter particles.

About a year after the big bang, the horizon surrounding any
point encompassed a mass of about $10^{12} M_\odot$, the
mass now in the dark matter halo of a large galaxy like the
Milky Way.  The temperature then was about a kilovolt.  We
define {\it cold} dark matter as particles that were moving
sluggishly, and {\it hot} dark matter as particles that were
still relativistic, at that time.  As Kim Griest and
Antonio Masiero discussed in their Varenna lectures, the
lightest superpartner particle (LSP neutralino) and the
axion remain the best motivated cold dark matter candidates,
although of course many other possibilities have been
suggested.

The three known neutrino species $\nu_e$, $\nu_\mu$, and
$\nu_\tau$ are the standard hot dark matter candidates.
Their contribution to the cosmological density today is
$$\Omega_\nu =  {\sum_i m(\nu_i) \over {94 h^2 \, {\rm eV}}}
$$
Since $\Omega_\nu < \Omega_0 \lsim 2$, each neutrino's mass
must be much less than a keV, so they were certainly
moving at relativistic speeds a year after the big bang.
Any of these neutrinos that has a cosmologically significant
mass ($\gsim 1$ eV) is therefore a hot dark matter particle.

If a horizon-sized region has slightly higher than average density at
this time, cold dark matter --- moving sluggishly --- will preserve
such a fluctuation.  But light neutrinos --- moving at nearly the speed of
light --- will damp such fluctuations by ``free streaming.'' For
example, two years after the big bang, the extra neutrinos will have
spread out over the now-larger horizon.  The smallest fluctuations
that will not suffer this fate are those that come into the horizon
when the neutrinos become nonrelativistic, i.e. when the temperature
drops below the neutrino mass.  In a universe in which most of the dark
matter is hot, primordial fluctuations will damp on all scales up to
superclusters (with mass $\sim 10^{16} M_\odot$), leading to a
sequence of cosmogony (cosmological structure formation) in which
galaxies form only after superclusters. But this is contrary to
observations, which show galaxies to be old but superclusters still
forming.  Indeed, with fluctuations on large scales consistent with
the COBE upper limit, standard HDM models (i.e.  with the dark
matter being mostly neutrinos, and a Zel'dovich spectrum of
Gaussian adiabatic fluctuations) cannot form a significant
number of galaxies by the present.  Thus most current
comparisons of cosmological models with observations have
focused on models in which most of the dark matter is cold.

The standard CDM model \cite{BFPR84} assumed a Zel'dovich
(i.e. $n_p=1$) spectrum of primordial Gaussian adiabatic
fluctuations with $\Omega=1$.  It had the great virtues of
simplicity and predictive power, since it had only one free
parameter, the amplitude or bias $b$.  Moreover, for a while
it even looked like it agreed with all available data, with
$b \approx 2.5$.  One early warning that all was not well
for CDM was the cosmic background dipole anisotropy,
indicating a large velocity of the local group with respect
to the cosmic background radiation rest frame, about 600 km
s$^{-1}$.  I confess that I and many other theorists did not
immediately
appreciate its possibly devastating impact.  However, as
evidence began to accumulate, starting in 1986, that
there were large-scale flows of galaxies with such
velocities \cite{Dress87}, it became clear that standard
CDM could fit
these large-scale galaxy peculiar velocities (i.e. motions
in addition to the general Hubble expansion) only for
$b\approx 1$.  Standard CDM had various problems for any
value of $b$; for example, the CDM matter correlation
function, and hence also the galaxy and cluster
correlations, are negative on scales larger than about $30
\hMpc$, while observations on these large scales show that
the cluster correlations are at least $\sim 3\sigma$
positive \cite{Olivier93}.
A low value of the bias parameter subsequently
also turned out to be required by the COBE DMR data, which
was first announced in April 1992.  But for such a small $b
\lsim 1$, CDM produces far too many clusters and predicts
small-scale galaxy velocities that are much too large
\cite{White93}.  Thus standard CDM does not look like a very
good match to the now-abundant observational data.  But it
did not miss by much: if the bias parameter $b$ is adjusted
to fit the COBE data, the fluctuation amplitude is too large on small
scales by perhaps a factor of $\sim 2-3$.

In the wake of the discovery of the existence of large-scale
galaxy peculiar velocities, I suggested that Jon Holtzman
(then a UCSC graduate student whose planned Ph.D. research
based on HST observations had been indefinitely postponed by
the Challenger explosion) improve the program that George
Blumenthal and I had written to do linear CDM calculations,
and use it to investigate a variety of models in which the
dark matter was mostly cold.  He ultimately worked out a
total of 94 such models, about half of them including some
hot dark matter, and (since this was the largest such suite
of interesting models all worked out the same way) his
thesis \cite{Holtz89} provided the basis for the COBE-DMR
interpretation paper \cite{Wright92}.  Meanwhile, in a
follow-up paper \cite{HP93}, we showed that of all these
CDM-like models the ones that best fit the available data
--- especially the cluster correlations --- were $\Omega=1$
Cold + Hot Dark Matter (CHDM),
and low-$\Omega$ Cold Dark Matter with a Cosmological
Constant ($\Lambda$CDM).
Since both of these models turned out to fit all available
data rather well when their fluctuation amplitudes were
normalized to COBE observations, they remain perhaps the
most popular models for galaxy formation and large scale
structure.  Moreover, since CHDM works best for $h \approx
0.5$ while $\Lambda$CDM works best for higher $h$, they will
serve nicely for this review as representatives of these two
opposing alternatives.

\section{Age of the Universe $t_0$}

The strongest lower limits for $t_0$ come from studies of
the stellar populations of globular clusters (GCs).
Standard estimates of the ages of the oldest GCs are 14-18
Gyr \cite{BolteHogan}, and a conservative lower limit on the
age of GCs is $13\pm2$ Gyr \cite{NewGC}, which is then a
lower limit on $t_0$. The main uncertainty in the GC age
estimates comes from the uncertain distance to the GCs: a
0.25 magnitude error in the distance modulus translates to a
22\% error in the derived cluster age \cite{Chaboyer94}.
Stellar mass loss is a recent idea for lowering the GC
$t_0$\cite{Shi94}, but observations constrain the reduction
in $t_0$ to be less than $\sim 1$ Gyr.  Allowing $\sim 1-2$
Gyr for galaxy and GC formation, we conclude that $t_0 \gsim
11$ Gyr from GCs, with $t_0 \approx 13$ Gyr a ``likely''
lower limit on $t_0$, obtained by pushing many but not all
the parameters to their limits.

The GC age estimates are of course based on standard stellar
evolution calculations.  But the solar neutrino problem
reminds us that we are not really sure that we understand
how even our nearest star operates; and the sun plays an
important role in calibrating stellar evolution, since it is
the only star whose age we know independently (from
radioactive dating of early solar system material).  What if
the GC age estimates are wrong for some unknown reason?

The only independent estimates of the age of the universe
come from cosmochronometry --- the chemical evolution of the
Galaxy --- and white dwarf cooling.  Cosmochronometry age
estimates are sensitive to a number of uncertain effects
such as the formation history of the disk and its stars, and
possible actinide destruction in stars \cite{Cosmochron}.
Age estimates also come from the cooling of white dwarfs in
the neighborhood of the sun.  The key observation is that
there is a lower limit to the luminosity and therefore also
the temperature of nearby white dwarfs; although dimmer ones
could have been seen, none have been found.  The only
plausible explanation is that the white dwarfs have not had
sufficient time to cool to lower temperatures, which
initially led to an estimate of $9.3\pm2$ Gyr for the age of
the Galactic disk \cite{Winget}.  Since there is evidence
that the stellar disk of our Galaxy is about 2 Gyr younger
than the oldest GCs \cite{VdB}, this in turn gave an
estimate of the age of the universe of $t_0 \sim 11\pm2$
Gyr.  More recent analyses \cite{Yuan} conclude that
sensitivity to disk star formation history, and to effects
on the white dwarf cooling rates due to C/O separation at
crystallization and possible presence of trace elements such
as $^{22}$Ne, allow a rather wide range of ages for the disk
of about $10\pm4$ Gyr. The latest determination of the white
dwarf luminosity function, using white dwarfs in proper
motion binaries, leads to a somewhat lower minimum
luminosity and therefore a somewhat higher estimate of the
age of the disk of $\sim 10.5^{+2.5}_{-1.5}$ Gyr; it follows
that $t_0\geq 11.5$ Gyr \cite{Wood96}.

\begin{figure}[htb]
\centering
\centerline{\epsfig{file=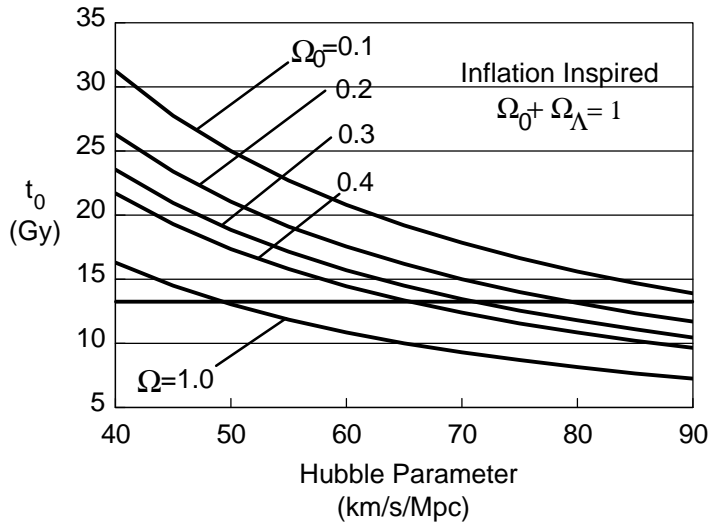,height=8cm}}
\caption{Age of the universe $t_0$ as a function of Hubble parameter
$H_0$ in inflation inspired models with $\Omega_0 + \Omega_\Lambda = 1$,
for several values of the present-epoch cosmological density
parameter $\Omega_0$.}
\end{figure}

Suppose that the GC stellar age estimates that $t_0 \gsim
13$ Gyr are right.  Fig.~1 shows that $t_0 > 13$ Gyr implies
that $H_0 \leq 50$ km s$^{-1}$ Mpc$^{-1}$ for $\Omega=1$,
and that $H_0 \leq 81$ km s$^{-1}$ Mpc$^{-1}$ even for
$\Omega_0 $ as small as 0.2 (in flat cosmologies with $\Omega_0
+ \Omega_\Lambda = 1$).

\section{Hubble Parameter $H_0$}

The Hubble parameter $H_0\equiv 100 h$ km s$^{-1}$ Mpc$^{-1}$
remains uncertain by about a factor of two: $0.4\lsim h
\lsim 1$.  Sandage has long contended that $h \approx 0.5$,
and he still concludes \cite{SandageRev,SandageTam} that the
latest data are consistent with $h=0.55\pm0.1$.  de
Vaucouleurs long contended that $h \approx 1$. A majority of
observers currently favor a value intermediate between these
two extremes (recent reviews include
\cite{Huchra92,Jacoby92,Fukugita93,KennFM}).

The Hubble parameter has been measured in two basic ways:
(A) Measuring the distance to some nearby galaxies,
typically by measuring the periods and luminosities of
Cepheid variables in them; and then using these ``calibrator
galaxies'' to set the zero point in any of the several
methods of measuring the relative distances to galaxies.
(B) Using fundamental physics to measure the distance to
some distant object directly, thereby avoiding at least some
of the uncertainties of the cosmic distance ladder
\cite{RR85}.  The difficulty with method (A) is that there
are so far only a handful of calibrator galaxies close
enough for Cepheids to be resolved in them.  However, the
success of the HST Cepheid measurement of the distance to
M100 \cite{Freedman94} shows that the HST Key Project on the
Extragalactic Distance Scale can significantly increase the
set of calibrator galaxies within a few years.  Adaptive
optics from the ground may also be able to contribute to
this effort, although I am not very impressed by the first
published result of this approach \cite{Pierce94}.  The
difficulty with method (B) is that in every case studied so
far, some aspect of the observed system or the underlying
physics remains somewhat uncertain.  It is nevertheless
remarkable that the results of several different methods of
type (B) are rather similar, and indeed not very far from
those of method (A).  This gives reason to hope for
convergence.

\subsection{(A) Relative Distance Methods}
One piece of good news is that the several methods of
measuring the relative distances to galaxies now mostly seem
to be consistent with each other \cite{Jacoby92,Fukugita93}.
These methods use either (1) ``standard candles'' or (2)
empirical relations between two measurable properties of a
galaxy, one distance-independent and the other
distance-dependent.  The old favorite standard candle is
Type Ia supernovae; a new one is the apparent maximum
luminosity of planetary nebulae \cite{Jacoby92}.  Sandage
\cite{SandageTam} and others \cite{vdBergh96} get low
values of $h\approx0.55$ from HST Cepheid distances to SN Ia
host galaxies, including the seven SNe Ia with well-observed
maxima that lie in six galaxies with HST Cepheid distances.
There are claims that taking account of an empirical
relationship between the SN Ia light curve shape and maximum
luminosity leads to higher $h$ \cite{Press94}, but Sandage
and Tammann counter that any such effect is small
\cite{SvsPhil,SandageTam} since they have not used those
``abnormal'' SNe Ia that would be significantly affected.
The old favorite empirical relation used as a relative
distance indicator is the Tully-Fisher relation between the
rotation velocity and luminosity of spiral galaxies (and the
related Faber-Jackson or $D_n - \sigma$ relation); a newer
one is based on the decrease in the fluctuations in
elliptical galaxy surface brightness on a given angular
scale as galaxies are seen at greater distances
\cite{Tonry}.

\subsection{(B) Fundamental Physics Approaches}
The fundamental physics approaches involve either Type Ia or
Type II supernovae, the Sunyaev-Zel'dovich (S-Z) effect, or
gravitational lensing.

The $^{56}$Ni radioactivity method for determining $H_0$
using Type Ia SN avoids the uncertainties of the distance
ladder by calculating the absolute luminosity of Type Ia
supernovae from first principles using a plausible but as
yet unproved physical model.  The first result obtained was
that $h=61\pm10$ \cite{Branch}; however, another study
\cite{Leibundgut} found that uncertainties in extinction
(i.e., light absorption) toward each supernova increases the
range of allowed $h$.  Demanding that the $^{56}$Ni
radioactivity method agree with an expanding photosphere
approach leads to $H_0=60^{+14}_{-11}$ \cite{Nugent}.
The expanding photosphere method compares the expansion rate
of the SN envelope measured by redshift with its size
increase inferred from its temperature and magnitude.  This
approach was first applied to Type II SN; the 1992 result
$h=0.6\pm0.1$ \cite{SKE92} was subsequently revised upward
by the same authors to $h=0.73\pm0.06\pm0.07$
\cite{Schmidt94}.  However, there are various complications
with the physics of the expanding envelope \cite{SNIIenv}.

The S-Z effect is the Compton scattering of microwave
background photons from the hot electrons in a foreground
galaxy cluster.  This can be used to measure $H_0$ since
properties of the cluster gas measured via the S-Z effect
and from X-ray observations have different dependences on
$H_0$.  The result from the first cluster for which
sufficiently detailed data was available, A665 (at
$z=0.182$), was $h=(0.4-0.5)\pm0.12$ \cite{Birk91};
combining this with data on A2218 ($z=0.171$) raises this
somewhat to $h=0.55\pm0.17$ \cite{Birk94}.  Early results
from the ASCA X-ray satellite gave $h=0.47\pm0.17$ for A665
($z=0.182$) and $h=0.41^{+0.15}_{-0.12}$ for CL0016+16
($z=0.545$) \cite{ASCA}.  A few S-Z results have been obtained using
millimeter-wave observations, and this promising method should
allow many more such measurements soon \cite{Lange}.
Corrections for the near-relativistic electron motions will
raise these estimates for $H_0$ a little \cite{Raphaeli},
but it seems clear that the S-Z results favor a smaller
value than many optical astronomers obtain.  However, since
the S-Z measurement of $H_0$ is affected by the orientation
of the cluster ellipticity with respect to the line of
sight, this will only become convincing with observations of
a significant number of additional clusters.  Fortunately,
this now appears to be possible within the next several
years.

Several quasars have been observed to have multiple images
separated by a few arc seconds; this phenomenon is
interpreted as arising from gravitational lensing of the
source quasar by a galaxy along the line of sight.  In the
first such system discovered, QSO 0957+561 ($z=1.41$), the
time delay $\Delta t$ between arrival at the earth of
variations in the quasar's luminosity in the two images has
been measured to be $409\pm23$ days \cite{Pelt94}, although
other authors found a value of $540\pm12$ days
\cite{Press92}.  The shorter $\Delta t$ has now been
confirmed by the observation of a sharp drop in Image A of
about 0.1 mag in late December 1994 \cite{ETurner95},
followed by a similar drop in Image B about 405-420 days later
(R. Schild and E.L. Turner, private communications).
Since $\Delta t \approx \theta^2 H_0^{-1}$,
this observation allows an estimate of the Hubble parameter,
with the results $h=0.50\pm0.17$ \cite{Rhee91}, or
$h=0.63\pm0.21$ ($h=0.42\pm0.14$) including (neglecting) dark
matter in the lensing galaxy \cite{Roberts91}, with
additional uncertainties associated with possible
microlensing and unknown matter distribution in the lensing
galaxy.  However, recent deep images have allowed mapping of
the gravitational potential of the cluster (at $z=0.36$) in
which the lensing galaxy lies using weak gravitational
lensing, which leads to the conclusion that $h \leq 0.70(1.1
{\rm yr}/\Delta t)$ \cite{Danle}. Also, detailed study of
the lensed QSO images (which include a jet) constrains the
lensing and implies $h= 0.82(1-\kappa)(1.1 {\rm yr}/\Delta
t)<0.82$, where the upper limit follows because the
convergence due to the cluster $\kappa>0$, or alternatively
$h=0.82(\sigma/322\kms)^2(1.1 {\rm yr}/\Delta t)$ without
uncertainty concerning the cluster if the one-dimensional
velocity dispersion $\sigma$ in the core of the giant
elliptical galaxy responsible for the lensing can be
measured \cite{Narayan}. Although the uncertainty in $H_0$
remains rather large, it is reassuring that this method
gives results consistent with the other determinations.  The
time-delay method is promising, and when delays are reliably
measured in several other multiple-image quasar systems,
that should lead to a reliable value for $h$.

\subsection{Correcting for Virgocentric Infall}
What about the recent HST Cepheid measurement of $H_0$,
giving $h\approx0.8$ \cite{Freedman94}?  This calculated
value is based on neither of the two methods (A) or (B)
above, and I do not regard it as being very reliable.
Instead this result is obtained by assuming that M100 is at
the core of the Virgo cluster, and dividing the sum of the
recession velocity of Virgo, about 1100 km s$^{-1}$, plus
the calculated ``infall velocity'' of the local group toward
Virgo, about 300 km s$^{-1}$, by the measured distance to
M100 of 17.1 Mpc.  (These recession and infall velocities
are both a little on the high side, compared to other values
one finds in the literature.)  Adding the ``infall
velocity'' is necessary in this method in order to correct
the Virgo recession velocity to what it would be were it
not for the gravitational attraction of Virgo for the Local
Group of galaxies, but the problem with this is that the net
motion of the Local Group with respect to Virgo is
undoubtedly affected by much besides the Virgo cluster ---
e.g., the ``Great Attractor.''  For example, in our CHDM
supercomputer simulations (which appear to be a rather
realistic match to observations) Anatoly Klypin and I have
found that galaxies and groups at about 20 Mpc from a
Virgo-sized cluster often have net outflowing rather than
infalling velocities.  Note that if the net ``infall'' of
M100 were smaller, or if M100 were in the foreground of the
Virgo cluster (in which case the actual distance to Virgo
would be larger than 17.1 Mpc), then the indicated $H_0$
would be smaller.

The authors of Ref.~\cite{Freedman94} gave an alternative
argument that avoids the ``infall velocity'' uncertainty:
the relative galaxy luminosities indicate that the Coma
cluster is about six times farther away than the Virgo
cluster, and peculiar motions of the Local Group and the
Coma cluster are relatively small corrections to the much larger
recession velocity of Coma; dividing the recession velocity
of the Coma cluster by six times the distance to M100 again
gives $H_0\approx80$.  However, this approach still assumes
that M100 is in the core rather than the foreground of the
Virgo cluster; and in deducing the relative distance of the
Coma and Virgo clusters it assumes that the galaxy
luminosity functions in each are comparable, which is
uncertain in view of the very different environments.  More
general arguments by the same authors \cite{Mould95} lead
them to conclude that $H_0=73\pm11$ regardless of where M100
lies in the Virgo cluster.  But Tammann et al.
\cite{SandageTam}, using all the available HST Cepheid
distances and their own complete sample of Virgo spirals,
conclude that $H_0\approx 54$.

\bigskip
To summarize, many observers, using mainly method (A), favor
a value $h\approx 0.6-0.8$ although Sandage's group and some
others continue to get $h\approx 0.5-0.6$, while the
methods I have grouped together as (B) typically lead to $h
\approx 0.4-0.7$.  The fact that the latter measurements are
mostly of more distant objects has suggested \cite{lowH}
that the local universe may actually be underdense and
therefore be expanding faster than is typical.  But in
reasonable models where structure forms from Gaussian
fluctuations via gravitational instability, it is extremely
unlikely that a sufficiently large region has a density
sufficiently smaller than average to make more than a rather
small difference in the measured value of $h$ \cite{Suto94}.
Moreover, the small dispersion in the corrected maximum
luminosity of distant SNe Ia found by the LBL Supernova
Cosmology Project \cite{KimLBL} compared to nearby SNe Ia
shows directly that the local and cosmological values of
$H_0$ are approximately equal.

There has been recent observational progress in both
methods (A) and (B), and I think it likely that the Hubble
parameter will be known reliably to 10\% within a few years.
But at present the uncertainty remains rather large
$0.4<h<0.8$, and we must keep an open mind.

\section{Cosmological Constant $\Lambda$, and $t_0$
Again}

Inflation is the only known solution to the horizon and
flatness problems and the avoidance of too many GUT
monopoles.  And inflation has the added bonus that with no
extra charge (except the perhaps implausibly fine-tuned
adjustment of the self-coupling of the inflaton field to be
adequately small), simple inflationary models predict a
near-Zel'dovich spectrum (i.e., with $n_p\approx1$) of
adiabatic Gaussian primordial fluctuations --- which seems
to be consistent with observations. All simple inflationary
models predict that the curvature constant $k$ is
vanishingly small, although inflationary models that are
extremely contrived (at least, to my mind) can be
constructed with negative curvature and therefore $\Omega_0
\ltwid 1$ without a cosmological constant \cite{RatraP}.  Thus
most authors who consider inflationary models impose the
condition $k=0$, or $\Omega_0 + \Omega_\Lambda =1$ where
$\Omega_\Lambda \equiv \Lambda/(3H_0^2)$.  This is what is
assumed in $\Lambda$CDM models, and it is what was assumed
in Fig.~1.  (I hope it has been clear from the foregoing
that I use $\Omega$ to refer only to the density of matter
and energy, not including the cosmological constant, whose
contribution in the $\Omega$ units is $\Omega_\Lambda$.)

I know of no one (except possibly Lev Kofman) who actually
finds the idea of a nonvanishing $\Lambda$ intrinsically
attractive.  There is no known physical reason why $\Lambda$
should be so small (from the viewpoint of particle physics),
though there is also no known reason why it should vanish.
The most unattractive features of $\Lambda \ne0$ cosmologies
are the fact that $\Lambda$ must become important only at
relatively low redshift --- why not much earlier or much
later? --- and also that $\Omega_\Lambda \gsim \Omega_0$
implies that the universe has recently entered an
inflationary epoch (with a de Sitter horizon comparable to
the present horizon).  The main motivations for $\Lambda >
0$ cosmologies are (1) reconciling inflation with
observations that seem to imply $\Omega_0 \ltwid 1$, and (2)
avoiding a contradiction between the lower limit $t_0 \gsim
13$ Gyr from globular clusters and $t_0=(2/3)H_0^{-1}=6.52
h^{-1}$ Gyr for the standard $\Omega=1$, $\Lambda=0$
Einstein-de Sitter cosmology, if it is really true that $h >
0.5$.

The cosmological effects of a cosmological constant are not
difficult to understand \cite{LLPR,LambdaARAA}.  With a
positive $\Lambda$, there is a repulsion of space by space.
In the early universe, the density of energy and matter is
far more important than $\Lambda$ on the r.h.s. of the
Friedmann equation.  But the average matter density
decreases as the universe expands, and at a rather low
redshift $z \sim 0.3$ the $\Lambda$ term finally becomes
dominant.  If it has been adjusted just right, $\Lambda$
can almost balance the attraction of the matter, and the
expansion nearly stops: for a long time, the scale factor $a
\equiv (1+z)^{-1}$ increases very slowly, although it
ultimately starts increasing exponentially as the universe
starts inflating under the influence of the increasingly
dominant $\Lambda$ term.  The existence of a period during
which expansion slows while the clock runs explains why
$t_0$ can be greater than for $\Lambda=0$, but this also
shows that there is a increased likelihood of finding
galaxies at the redshift interval when the expansion slowed,
and a correspondingly increased opportunity for lensing of
quasars (which mostly lie at higher redshift $z \gsim 2$) by
these galaxies.

The frequency of such lensed quasars is about what would be
expected in a standard $\Omega=1$, $\Lambda=0$ cosmology,
so this data sets fairly stringent upper limits:
$\Omega_\Lambda \leq 0.70$ at 90\% C.L.
\cite{MaozRix,Kochanek94}, with more recent data
giving even tighter constraints: $\Omega_\Lambda < 0.66$ at
95\% confidence \cite{Kochanek95}.  This limit could
perhaps be weakened if there were (a) significant extinction
by dust in the E/S0 galaxies responsible for the lensing or
(b) rapid evolution of these galaxies, but there is much
evidence that these galaxies have little dust and have
evolved only passively for $z\lsim1$ \cite{SCarlberg}.

A weaker but independent constraint comes from the cosmic
background radiation data.  In standard $\Omega=1$ models,
the quantity $\ell(\ell+1)C_\ell$ (where
$C_\ell=<a_{\ell,m}^2>_m$ is the average of squared
coefficients of the spherical harmonic expansion of the CMB
data) is predicted to be roughly constant for $2\leq \ell
\lsim 10$ (with an increase for
higher multiples toward the Doppler peak at $\ell \sim
200$), while in models with $\Lambda>0$
$\ell(\ell+1)C_\ell$ is predicted to dip before rising
toward the Doppler peak.  Comparison with the two-year COBE
data, in which such a dip is not seen, implies that
$\Omega_\Lambda \leq 0.78$ at the 90\% C.L. \cite{BunnS94}.

Yet another constraint comes from number counts of bright
E/S0 galaxies in HST images \cite{Driver96}, since these
galaxies appear to have evolved rather little since
$z\sim1$ \cite{SCarlberg}.  The number counts are just as expected in the
$\Omega=1$, $\Lambda=0$ Einstein-de Sitter cosmology.  Even
allowing for uncertainties due to evolution and merging of
these galaxies, this data would allow $\Omega_\Lambda$ as
large as 0.8 in flat cosmologies only in the unlikely event
that half the Sa galaxies in the deep HST images were
misclassified as E/S0. This approach may be very promising
for the future, as the available deep HST image data and our
understanding of galaxy evolution both increase.

A model-dependent constraint comes from a detailed
simulation of $\Lambda$CDM \cite{KPH96}: a COBE-normalized
model with $\Omega_0=0.3$ and $h=0.7$ has far too much power
on small scales to be consistent with observations, even
allowing for seemingly unrealistic antibiasing of galaxies
with respect to dark matter.  For $\Lambda$CDM models, the
only solution appears to be raising $\Omega_0$, lowering
$H_0$, and tilting the spectrum ($n_{p}<1$) \cite{KPH96},
though of course one could alternatively modify the
primordial power spectrum in other ways.

Fig.~1 shows that with $\Omega_\Lambda \leq 0.7$, the
cosmological constant does not lead to a very large increase
in $t_0$ compared to the Einstein-de Sitter case, although
it may still be enough to be significant.  For example, the
constraint that $t_0 \ge 13$ Gyr requires $h \leq 0.5$ for
$\Omega=1$ and $\Lambda=0$, but this becomes $h \leq 0.70$
for flat cosmologies with $\Omega_\Lambda \leq 0.66$.

\section{Measuring $\Omega_0$}

\subsection{Very Large Scale Measurements}

Although it would be desirable to measure $\Omega_0$ and
$\Lambda$ through their effects on the large-scale geometry
of space-time, this has proved difficult in practice since
it requires comparing objects at higher and lower redshift,
and it is hard to separate the effects of the evolution of
the objects from those of the evolution of the universe.
For example, in ``redshift-volume'' tests involving number
counts of galaxies per redshift interval, how can we tell
whether the galaxies at redshift $z\sim 1$ correspond to
those at $z \sim 0$?  Several galaxies at higher redshift
might have merged, and galaxies might have formed or changed
luminosity at lower redshift.  Eventually, with extensive
surveys of galaxy properties as a function of redshift using
the largest telescopes such as Keck, it should be possible
to perform these classical cosmological tests at least on a
particular class of galaxies --- that is one of the goals of
the Keck DEEP project.

At present, perhaps the most promising technique involves
searching for Type Ia supernovae (SNe Ia) at high-redshift,
since these are the brightest supernovae and the spread in
their intrinsic brightness appears to be relatively small.
Saul Perlmutter, Gerson Goldhaber, and collaborators have
recently demonstrated the feasibility of finding significant
numbers of such supernovae \cite{Perlmutter95}, but a
dedicated campaign of follow-up observations of each one
will be required in order to measure $\Omega_0$ by
determining how the apparent brightness of the supernovae
depends on their redshift.  This is therefore a demanding
project.  It initially appeared that $\sim 100$ high
redshift SNe Ia would be required to achieve a 10\%
measurement of $q_0=\Omega_0/2 - \Omega_\Lambda$.  However,
using the correlation mentioned earlier between the absolute
luminosity of a SN Ia and the shape of its light curve
(e.g., slower decline correlates with higher peak
luminosity), it now appears possible to reduce the number of
SN Ia required.  The Perlmutter group has now
analyzed seven high redshift SN Ia by this method, with a
tentative result that $q_0 \sim 0.5$ and in any case $q_0
\gsim 0$ \cite{Perlmutter96}.  In November 1995 they discovered
an additional 11 high-redshift SN Ia, and they have just
discovered an additional seven.  Preliminary results from
the next 11 SN Ia are consistent with $q_0 \sim 0.5$,
and there are no peculiar or anomalous ones in the dataset
(Perlmutter, private communication).  Two other groups,
collaborations from ESO and MSSSO/CfA/CTIO, are also
searching for high-redshift supernovae to measure $q_0$, and
have found at least a few.  There has also been recent
progress understanding the physical origin of the SN Ia
luminosity-light curve correlation.  At the present rate of
progress, a reliable answer may be available within perhaps
a year or so if a consensus emerges from these efforts.

\subsection{Large-scale Measurements}
$\Omega_0$ has been measured with some precision on a scale
of about $\sim 50 \hMpc$, using the data on peculiar
velocities of galaxies, and on a somewhat larger scale using
redshift surveys based on the IRAS galaxy catalog.  Since
the results of all such measurements to date have recently
been reviewed in detail \cite{DekelARAA}, I will only
comment briefly on them.  The analyses such as ``POTENT''
that try to recover the scalar velocity potential from the
galaxy peculiar velocities are looking increasingly
reliable, since they reproduce the observed large scale
distribution of galaxies -- that is, many galaxies are found
where the converging velocities indicate that there is a lot
of matter, and there are voids in the galaxy distribution
where the diverging velocities indicate that the density is
lower than average.  The comparison of the IRAS redshift
surveys with POTENT and related analyses typically give
fairly large values for the parameter $\beta_I \equiv
\Omega_0^{0.6}/b_I$ (where $b_I$ is the biasing parameter
for IRAS galaxies), corresponding to $0.3 \lsim \Omega_0
\lsim 3$ (for an assumed $b_I=1.15$).  It is not clear
whether it will be possible to reduce the spread in these
values significantly in the near future --- probably both
additional data and a better understanding of systematic and
statistical effects will be required.

A particularly simple way to deduce a lower limit on
$\Omega_0$ from the POTENT peculiar velocity data has recently
been proposed \cite{DekRees}, based on the fact that
high-velocity outflows from voids are not expected in
low-$\Omega$ models.  Data on just one void indicates that
$\Omega_0 \ge 0.3$ at the 97\% C.L.  This argument is
independent of assumptions about $\Lambda$ or galaxy
formation, but of course it does depend on the success of
POTENT in recovering the peculiar velocities of galaxies.

However, for the particular cosmological models that I am
focusing on in this review --- CHDM and $\Lambda$CDM ---
stronger constraints are available. This is because these
models, in common with almost all CDM variants, assume that
the probability distribution function (PDF) of the
primordial fluctuations was Gaussian.  Evolution from a
Gaussian initial PDF to the non-Gaussian mass distribution
observed today requires considerable gravitational
nonlinearity, i.e. large $\Omega$.  The PDF deduced by
POTENT from observed velocities (i.e., the PDF of the mass,
if the POTENT reconstruction is reliable) is far from
Gaussian today, with a long positive-fluctuation tail.  It
agrees with a Gaussian initial PDF if and only if $\Omega$
is about unity or larger: $\Omega_0 <1$ is rejected at the
$2\sigma$ level, and $\Omega_0 \leq 0.3$ is ruled out at
$\ge  4\sigma$ \cite{NusDek}.

\subsection{Measurements on Scales of a Few Mpc}
On smaller length scales, there are many measurements that
are consistent with a smaller value of $\Omega_0$
\cite{Peebles}.  For example, the cosmic virial theorem
gives $\Omega(\sim 1 h^{-1}\,{\rm Mpc}) \approx 0.15
[\sigma(1 h^{-1} \, {\rm Mpc}) / (300 \, {\rm km} \,
{\rm s}^{-1})]^2$, where $\sigma(1 h^{-1} \, {\rm Mpc})$ here
represents the relative velocity dispersion of galaxy pairs
at a separation of $1 h^{-1} \, {\rm Mpc}$.  Although the
classic paper \cite{DavisP} which first measured $\sigma(1
h^{-1} \, {\rm Mpc})$ using a large redshift survey (CfA1)
got a value of 340 km s$^{-1}$, this result is now known to
be in error since the entire core of the Virgo cluster was
inadvertently omitted \cite{Somerville}; if Virgo is
included, the result is $\sim 500-600$ km s$^{-1}$
\cite{Moetal,Somerville}, corresponding to $\Omega(\sim 1
h^{-1}\,{\rm Mpc}) \approx 0.4-0.6$.  Various redshift
surveys give a wide range of values for $\sigma(1 h^{-1} \,
{\rm Mpc}) \sim 300-750$ km s$^{-1}$, with the most salient
feature being the presence or absence of rich clusters of
galaxies; for example, the IRAS galaxies, which are not
found in clusters, have $\sigma(1 h^{-1} \, {\rm Mpc})
\approx 320$ km s$^{-1}$ \cite{Fisher94}, while the northern
CfA2 sample, with several rich clusters, has much larger
$\sigma$ than the SSRS2 sample, with only a few relatively
poor clusters \cite{Marzke,Somerville,Somerville2}.  It is evident that the
$\sigma(1 h^{-1} \, {\rm Mpc})$ statistic is not a very
robust one.

A standard method for estimating $\Omega$ on scales of a
few Mpc is based on applying virial estimates to groups and
clusters of galaxies to try to deduce the total mass of the
galaxies including their dark matter halos from the
velocities and radii of the groups; roughly, $G M \sim r v^2$.
(What one actually does is to assume that all galaxies have
the same mass-to-light ratio $M/L$, given by the median
$M/L$ of the groups, and integrate over the luminosity
function to get the mass density \cite{KOS79,HuchraG82,Ramella89}.)
The typical result is that $\Omega(\sim 1 \hMpc)
\sim 0.1-0.2$.  However, such estimates are at best lower
limits, since they can only include the mass within the
region where the galaxies in each group can act as test
particles.  In CHDM simulations, my colleagues and I
\cite{NKP95} have found that the effective radius of the
dark matter distribution associated with galaxy groups is
typically 2-3 times larger than that of the galaxy
distribution.  Moreover, we find a velocity biasing
\cite{CarlCouch} factor in CHDM groups $b_v^{grp} \equiv
v_{\rm gal, rms}/v_{\rm DM, rms} \approx 0.75$, whose
inverse squared enters in the $\Omega$ estimate.  Finally,
we find that groups and clusters are typically elongated, so
only part of the mass is included in spherical estimators.
These factors explain how it can be that our $\Omega=1$ CHDM
simulations produce group velocities that are fully
consistent with those of observed groups, even with
statistical tests such as the median rms group velocity vs.
the fraction of galaxies grouped \cite{NKP94,NKP95}.  This
emphasizes the point that local estimates of $\Omega$ are at
best lower limits on its true value.

Another approach to estimating $\Omega$ from information on
relatively small scales has been pioneered by Peebles
\cite{PeebAction}.  It is based on using the least action
principle (LAP) to reconstruct the trajectories of the Local
Group galaxies, and the assumption that the mass is
concentrated around the galaxies.  This is a reasonable
assumption in a low-$\Omega$ universe, but it is not at all
what must occur in an $\Omega=1$ universe where most of the
mass must lie between the galaxies.  Although comparison
with $\Omega=1$ N-body simulations showed that the LAP often
succeeds in qualitatively reconstructing the trajectories,
the mass is systematically underestimated by a large factor
by the LAP method \cite{BranCarl}.  Unexpectedly, a
different study \cite{DunnLaf} found that the LAP method
underestimates $\Omega$ by a factor of 4-5 even in an
$\Omega_0=0.2$ simulation; the authors say that this
discrepancy is due to the LAP neglecting the effect of
``orphans'' --- dark matter particles that are not members
of any halo. Shaya, Peebles, and Tully \cite{TullyPS} have
recently attempted to apply the LAP to galaxies in the local
supercluster, again getting low $\Omega_0$.  The LAP
approach should be more reliable on this larger scale, but
the method still must be calibrated on N-body simulations of
both high- and low-$\Omega_0$ models before its biases can
be quantified.

\subsection{Estimates on Galaxy Halo Scales}
Recent work by Zaritsky and White \cite{Zaritsky} and
collaborators has shown that spiral galaxies have massive
halos.  A classic paper by Little and Tremaine
\cite{LitTre87} argued that the available data on the Milky
Way satellite galaxies required that the Galaxy's halo
terminate at about 50 kpc, with a total mass of only about
$2.5 \times 10^{11} M_\odot$.  But by 1991, new data on local
satellite galaxies, especially Leo I, became available, and
the Little-Tremaine estimator increased to $1.25 \times
10^{12} M_\odot$.
A recent, detailed study finds a mass inside 50 kpc of
$(5.4\pm1.3)\times 10^{11} M_\odot$ \cite{KochanMWy}.
Zaritsky and collaborators have collected
data on satellites of isolated spiral galaxies, and conclude
that the fact that the relative velocities do not fall off
out to a separation of at least 200 kpc shows that massive
halos are the norm.  The typical rotation velocity of $\sim
200-250$ km  s$^{-1}$ implies a mass within 200 kpc of
$2\times10^{12} M_\odot$.  A careful analysis taking into
account selection effects and satellite orbit uncertainties
concluded that the indicated value of $\Omega_0$ exceeds 0.13
at 90\% confidence, with preferred values exceeding 0.3
\cite{Zaritsky}. Newer data suggesting that relative
velocities do not fall off out to a separation of at least
300 kpc will raise these $\Omega_0$ estimates
\cite{ZaritskyPC}.

However, if galaxy dark matter halos are really so extended
and massive, that would imply that when such galaxies
collide, the resulting tidal tails of debris cannot be flung
very far.  Therefore, the observed merging galaxies
with extended tidal tails such as NGC 4038/39 (the Antennae)
and NGC 7252 probably have halo:(disk+bulge) mass ratios
less than 10:1 \cite{Dubinski96}, unless the stellar tails
are perhaps made during the collision process from gas that
was initially far from the central galaxies (J. Ostriker,
private communication, 1996); the latter possibility can be
checked by determining the ages of the stars in these tails.

A direct way of measuring the mass and spatial extent of
many galaxy dark matter halos is to look for the small
distortions of distant galaxy images due to gravitational
lensing by foreground galaxies. This technique was pioneered
by Tyson et al. \cite{Tyson84}. Though the results were
inconclusive, powerful constraints could perhaps be obtained
from deep HST images or ground-based images with excellent
seeing.  Such fields would also be useful for measuring the
correlated distortions of galaxy images from large-scale
structure by weak gravitational lensing; although a pilot
project \cite{MouldBland} detected only a marginal signal,
a reanalysis detected a significant signal suggesting that
$\Omega_0 \sigma_8 \sim 1$ \cite{Villum95}.  Several groups
are planning major projects of this sort.

\section{Clusters}

\subsection{Cluster Baryons vs. Big Bang Nucleo\-synthesis}
A recent review \cite{Copi95} of Big Bang Nucleosynthesis (BBN) and
observations indicating primordial abundances of the light isotopes
concludes that $0.009h^{-2} \leq \Omega_b \leq 0.02h^{-2}$ for
concordance with all the abundances, and $0.006h^{-2} \leq \Omega_b
\leq 0.03h^{-2}$ if only deuterium is used.  For $h=0.5$, the
corresponding upper limits on $\Omega_b$ are 0.08 and 0.12,
respectively.  The recent observations \cite{Songaila} of a possible
deuterium line in a hydrogen cloud at redshift $z=3.32$ indicating a
deuterium abundance of $\sim 2\times 10^{-4}$ (and therefore $\Omega_b
\leq 0.006h^{-2}$) are contradicted by similar observations by Tytler
and collaborators \cite{Tytler} in systems at $z=3.57$ and $z=2.504$
but with a deuterium abundance
about ten times lower, consistent with solar system measurements of D
and $^3$He and implying $\Omega_b h^2=0.024\pm0.05$, or $\Omega_b$
in the range 0.08-0.11 for $h=0.5$.  (If these
represent the true D/H, then the earlier observations \cite{Songaila}
were most probably of a Ly$\alpha$ forest line.  However, Rugers and
Hogan \cite{Hogan95} argue that the width of their $z=3.32$ absorption
features is better fit by deuterium, although they admit that only a
statistical sample of absorbers will settle the issue.  There is a new
possible detection of D at $z=4.672$ in the absorption spectrum of QSO
BR1202-0725 \cite{Wampler} and at $Z=3.086$ toward Q 0420-388
\cite{Carswell96}.  But Tytler \cite{Tytler} argues that the two
systems he and his colleagues have analyzed are much more convincing
as real detections of deuterium, and it is surely significant that
they measure the same D/H in both systems.)

White et al. \cite{WhiteClus} have emphasized that recent
X-ray observations of clusters, especially Coma, show that
the abundance of baryons, mostly in the form of gas (which
typically amounts to several times the total mass of the
cluster galaxies), is about 20\% if $h$ is as low as
0.5. For the Coma cluster they find that the baryon fraction
within the Abell radius is
$$ f_b \equiv {M_b \over M_{tot}} \geq 0.009+0.050h^{-3/2},
$$
where the first term comes from the galaxies and the second
from gas.  If clusters are a fair
sample of both baryons and dark matter, as they are expected
to be based on simulations, then this is 2-3 times the
amount of baryonic mass expected on the basis of BBN in an
$\Omega=1$, $h\approx 0.5$ universe, though it is just what
one would expect in a universe with $\Omega_0 \approx 0.3$.
The fair sample hypothesis implies that
$$ \Omega_0 = {\Omega_b \over f_b} = 0.3 \left({\Omega_b \over
0.06}\right) \left({0.2 \over f_b}\right).
$$
A recent review of gas in a sample of clusters \cite{Fabian}
finds that the baryon mass fraction within about 1 Mpc lies
between 10 and 22\% (for $h=0.5$; the limits scale as $h^{-3/2}$),
and argues that it is unlikely that
(a) the gas could be clumped enough to lead to significant
overestimates of the total gas mass --- the main escape
route considered in \cite{WhiteClus} (cf. also \cite{ClusBar}).
The gas mass would also be overestimated if large tangled magnetic fields
provide a significant part of the pressure in the central regions of
some clusters \cite{LoebMao}; this can be checked by observation of
Faraday rotation of sources behind clusters \cite{Kronberg}.
If $\Omega=1$, the
alternatives are then either (b) that clusters have more mass
than virial estimates based on the cluster galaxy velocities
or estimates based on hydrostatic equilibrium \cite{BBlanchard}
of the gas at
the measured X-ray temperature (which is surprising since
they agree \cite{BLubin}),
or (c) that the BBN upper limit on $\Omega_b$
is wrong.  It is interesting that there are indications from
weak lensing \cite{Kaiser} that at least some clusters may 
actually have extended halos of dark matter --- something 
that is expected to a greater extent if the dark matter is a 
mixture of cold and hot components, since the hot component 
clusters less than the cold \cite{NKP95,BHNPK,KofmanKP}.  If 
so, the number density of clusters as a function of mass is 
higher than usually estimated, which has interesting 
cosmological implications (e.g. $\sigma_8$ is higher than 
usually estimated).  It is of course possible that the 
solution is some combination of alternatives (a), (b), and 
(c).  If none of the alternatives is right, then the only 
conclusion left is that $\Omega_0 \approx 0.3$.  The cluster 
baryon problem is clearly an issue that deserves very 
careful examination. 

\subsection{Cluster Morphology}
Richstone, Loeb, and Turner \cite{Richstone92} showed that
clusters are expected to be evolved --- i.e. rather
spherical and featureless --- in low-$\Omega$ cosmologies,
in which structures form at relatively high redshift, and
that clusters should be more irregular in $\Omega=1$
cosmologies, where they have formed relatively recently and
are still undergoing significant merger activity.  There are
very few known clusters that seem to be highly evolved and
relaxed, and many which are irregular --- some of which are
obviously undergoing mergers now or have recently done so
(see e.g. \cite{Burns}).  This disfavors low-$\Omega$
models, but it remains to be seen just how low.  Recent
papers have addressed this.  In one \cite{Mohr95} a total of
24 CDM simulations with $\Omega=1$ or 0.2, the latter with
$\Omega_\Lambda=0$ or 0.8, were compared with data on a
sample of 57 clusters.  The conclusion was that clusters
with the observed range of X-ray morphologies are very
unlikely in the low-$\Omega$ cosmologies.  However, these
simulations have been criticized because the $\Omega_0=0.2$
ones included rather a large amount of ordinary matter:
$\Omega_b=0.1$.  (This is unrealistic both because $h
\approx 0.8$ provides the best fit for $\Omega_0=0.2$ CDM, but
then the standard BBN upper limit is $\Omega_b<0.02h^{-
2}=0.03$; and also because observed clusters have a gas
fraction of $\sim 0.15(h/0.5)^{-3/2}$.)  Another study
\cite{Jing95} using dissipationless simulations and not
comparing directly to observational data found that
$\Lambda$CDM with $\Omega_0=0.3$ and $h=0.75$ produced
clusters with some substructure, perhaps enough to be
observationally acceptable.  Clearly, this important issue
deserves study with higher resolution hydrodynamic
simulations, with a range of assumed $\Omega_b$, and
possibly including at least some of the additional physics
associated with the galaxies which must produce the
metallicity observed in clusters, and perhaps some of the
heat as well.  Better statistics for comparing simulations
to data may also be useful \cite{Tsai}.

\subsection{Cluster Evolution}
There is evidence for strong evolution of clusters at
relatively low redshift, both in their X-ray properties
\cite{Gioia} and in the properties of their galaxies.
In particular, there is a strong increase in the fraction of
blue galaxies with increasing redshift (the
``Butcher-Oemler effect''), which may be difficult to
explain in a low-density universe \cite{Kauffmann94}.
Field galaxies do not appear to show such strong evolution;
indeed, a recent study concludes that over the redshift
range $0.2 \leq z \leq 1.0$ there is no significant evolution
in the number density of ``normal'' galaxies
\cite{Steidel94}.  This is compatible with the predictions
of CHDM with two neutrinos sharing a total mass of about 5
eV \cite{Klypin95} (see below).

\section{Early Structure Formation}

In linear theory, adiabatic density fluctuations grow
linearly with the scale factor in an $\Omega=1$ universe,
but more slowly if $\Omega<1$ with or without a cosmological
constant \cite{Peebles}.  As a result, if fluctuations of a
certain size in an $\Omega=1$ and an $\Omega_0=0.3$ theory
are equal in amplitude at the present epoch ($z=0$), then at
higher redshift the fluctuations in the low-$\Omega$ model
had higher amplitude.  Thus, structures typically form
earlier in low-$\Omega$ models than in $\Omega=1$ models.

Since quasars are seen at the highest redshifts, they have
been used to try to constrain $\Omega=1$ theories,
especially CHDM which because of the hot component has
additional suppression of small-scale fluctuations that
are presumably required to make early structure (e.g.,
\cite{Haehnelt}).  The difficulty is that dissipationless
simulations predict the number density of halos of a given
mass as a function of redshift, but not enough is known
about the nature of quasars --- for example, the mass of the
host galaxy --- to allow a simple prediction of the number
of quasars as a function of redshift in any given
cosmological model.  A recent study \cite{Katz94} concludes
that very efficient cooling of the gas in early structures,
and angular momentum transfer from it to the dark halo,
allows for formation of {\it at least} the observed number
of quasars even in models where most galaxy formation occurs
late.

Observers are now beginning to see significant numbers of 
what appear to be the central regions of galaxies in an 
early stage of their formation at redshifts $z=3-3.5$ 
\cite{Steidel96} --- although, as with quasars, a danger in 
using systems observed by emission is that they may not be 
typical.  As additional observations clarify the nature of 
these objects, they can perhaps be used to constrain 
cosmological parameters and models. 

Another sort of high redshift object which may hold more 
promise for constraining theories is damped Lyman $\alpha$ 
systems (DLAS).  DLAS are dense clouds of neutral hydrogen, 
generally thought to be protogalactic disks, which are 
observed as wide absorption features in quasar spectra 
\cite{Wolfe}.  They are relatively common, seen in roughly a 
third of all quasar spectra, so statistical inferences about 
DLAS are possible.  At the highest redshift for which data 
is published, $z=3-3.4$, the density of neutral gas in such 
systems in units of critical density is $\Omega_{gas} 
\approx 0.6$\%, comparable to the total density of visible 
matter in the universe today \cite{LWT}.  Several recent 
papers \cite{MoME} pointed out that the CHDM model with 
$\Omega_\nu=0.3$ could not produce such a high 
$\Omega_{gas}$.  However, my colleagues and I showed that 
CHDM with $\Omega_\nu=0.2$ could do so \cite{KBHP}, since 
the power spectrum on small scales is a very sensitive 
function of the total neutrino mass in CHDM models.  This 
theory makes two crucial predictions \cite{KBHP}: 
$\Omega_{gas}$ must fall off at higher redshifts $z$, and 
the DLAS at $z \gsim 3$ correspond to systems of internal 
rotation velocity or velocity dispersion less than about 100 
km s$^{-1}$ (this can be measured from the Doppler widths of 
the metal line systems associated with the DLAS).  
Preliminary reports regarding the amount of neutral hydrogen 
in such systems deduced from the latest data at redshifts 
above 3.5 appear to be consistent with these predictions 
\cite{Storrie}.  But a possible problem is the large 
velocity widths of the metal line systems associated with 
the highest-redshift DLAS yet reported \cite{LLu}, at 
$z=4.4$; if these actually indicate that a massive disk 
galaxy is already formed at such a high redshift, and if 
discovery of other such systems shows that they are not 
rare, that would certainly disfavor CHDM and other 
$\Omega=1$ theories with relatively little power on small 
scales.  Other interpretations of such data which would not 
cause such problems for theories like CHDM are perhaps more 
plausible, though \cite{Haehnelt96}. 

One of the best ways of probing early structure formation
would be to look at the main light output of the stars of
the earliest galaxies, which is redshifted by the expansion
of the universe to wavelengths beyond about 5 microns today.
Unfortunately, it is not possible to make such observations
with existing telescopes; since the atmosphere blocks
almost all such infrared radiation, what is required is a
large infrared telescope in space.  The Space Infrared
Telescope Facility (SIRTF) has long been a high priority,
and it would be great to have access to the data such an
instrument would produce.  In the meantime, an alternative
method is to look
for the starlight from the earliest stars as extragalactic
background infrared light (EBL).  Although it is difficult
to see this background light directly because our
Galaxy is so bright in the near infrared, it may be possible
to detect it indirectly through its absorption of TeV gamma
rays (via the process $\gamma \,\gamma \to e^+ \, e^-$). Of
the more than twenty active galactic nuclei (AGNs) that have 
been seen at $\sim10$ GeV by the EGRET detector on the 
Compton Gamma Ray Observatory, only two of the nearest, 
Mk421 and Mk501, have also been clearly detected in TeV 
gamma rays by the Whipple Atmospheric Cerenkov Telescope 
\cite{Whipple96}.  Absorption of $\sim$ TeV gamma rays  from  
(AGNs) at redshifts $z\sim 0.2$ has been shown to be a 
sensitive probe of the EBL and thus of the era of galaxy 
formation \cite{MacMinnP}. 

\section{Neutrino mass}

There are several experiments which suggest that neutrinos
have mass. In particular, the recent announcement of the
observation of $\bar\nu_\mu \to \bar\nu_e$ oscillations at
the Liquid Scintillator Neutrino Detector (LSND) experiment
at Los Alamos suggests that $\delta m^2 \equiv |m(\nu_\mu)^2
- m(\nu_e)^2| \approx 6$ eV$^2$ \cite{CaldwellStock}, and
the observation of the angular dependence of the atmospheric
muon neutrino deficit at Kamiokande \cite{Fukuda} suggests
$\nu_\mu \to \nu_\tau$ oscillations are occurring with an
oscillation length comparable to the depth of the
atmosphere, which requires that the muon and tau neutrinos
have approximately the same mass.  If, for example,
$m(\nu_e) \ll m(\nu_\mu)$, then this means that $m(\nu_\mu)
\approx m(\nu_\tau) \approx 2.4$ eV \cite{PHKC,FPQ}.  Clearly,
discovery of neutrino mass in the few eV range favors CHDM;
and, as I mentioned above, this total neutrino mass of about
5 eV is just what seems to be necessary to fit the large
scale structure observations \cite{KBHP}.  Dividing the mass
between two neutrino species results in somewhat lower 
fluctuation amplitude on the scale of clusters of galaxies 
because of the longer neutrino free streaming length, which 
improves agreement between CHDM normalized to COBE and 
observations of cluster abundance \cite{PHKC}. 

Of course, one cannot prove a theory since contrary evidence 
may always turn up.  But one can certainly disprove 
theories.  The minimum neutrino mass required by the 
preliminary LSND result \cite{CaldwellStock} $\delta m^2 = 
6$ eV$^2$ is 2.4 eV.  This is too much hot dark matter to 
permit significant structure formation in a low-$\Omega$ 
universe; for example, in a $\Lambda$CHDM model with 
$\Omega_0=0.3$, the cluster number density is more than two 
orders of magnitude lower than observations indicate 
\cite{PHKC}.  Thus if this preliminary LSND result is 
correct, it implies a strong lower limit on $\Omega_0$, and 
a corresponding upper bound on $\Lambda$, in $\Lambda$CDM 
models that include light neutrinos. 

\section{Conclusions}

The main issue that I have tried to address is the value of 
the cosmological density parameter $\Omega$.  Strong 
arguments can be made for $\Omega_0 \approx 0.3$ (and models 
such as $\Lambda$CDM) or for $\Omega=1$ (for which the best 
class of models that I know about is CHDM), but it is too 
early to tell for sure which is right. 

The evidence would favor a small $\Omega_0 \approx 0.3$ if (1) the
Hubble parameter actually has the high value $H_0 \approx 75$ favored
by many observers, and the age of the universe $t_0 \geq 13$ Gyr; or
(2) the baryonic fraction $f_b=M_b/M_{tot}$ in clusters is actually
$\sim 15$\%, about 3 times larger than expected for standard Big Bang
Nucleosynthesis in an $\Omega=1$ universe. This assumes that standard
BBN is actually right in predicting that the density of ordinary
matter $\Omega_b$ lies in the range $0.009 \leq \Omega_b h^2 \leq
0.02$.  High-resolution high-redshift spectra are now providing
important new data on primordial abundances of the light isotopes that
should clarify the reliability of the BBN limits on $\Omega_b$.  If
the systematic errors in the $^4$He data are larger than currently
estimated, then it may be wiser to use the deuterium upper limit
$\Omega_b h^2 \leq 0.03$, which is also consistent with the value
$\Omega_b h^2\approx 0.024$ indicated by the only clear deuterium
detection at high redshift, with the same D/H$\approx 2.4\times
10^{-5}$ observed in two different low-metallicity quasar absorption
systems \cite{Tytler}; this considerably lessens the discrepancy
between $f_b$ and $\Omega_b$.  Another important constraint on
$\Omega_b$ will come from the new data on small angle CMB anisotropies
--- in particular, the height of the first Doppler peak \cite{DGS95},
with the latest data consistent with low $h\approx 0.5$ and high
$\Omega_b \approx 0.1$.

The evidence would favor $\Omega=1$ if (1) the POTENT 
analysis of galaxy peculiar velocity data is right, in 
particular regarding outflows from voids or the inability to 
obtain the present-epoch non-Gaussian density distribution 
from Gaussian initial fluctuations in a low-$\Omega$ 
universe; or (2) the preliminary report from LSND indicating 
a neutrino mass $\geq 2.4$ eV is right, since that would be 
too much hot dark matter to allow significant structure 
formation in a low-$\Omega$ $\Lambda$CDM model. 

The statistics of gravitational lensing of quasars is 
incompatible with large cosmological constant $\Lambda$ and 
low cosmological density $\Omega_0$.  Discrimination between 
models may improve as additional examples of lensed quasars 
are searched for in large surveys such as the Sloan Digital
Sky Survey. 

It now appears to be possible to measure the deceleration 
parameter $q_0=\Omega_0/2-\Omega_\Lambda$ on very large 
scales using the objects that may be the best bright 
standard candles: high-redshift Type Ia supernovae. It is 
very encouraging that the Perlmutter group 
\cite{Perlmutter96} now has discovered $\sim25$ 
high-redshift Type Ia supernovae, that other groups are also 
succeeding in finding such supernovae, and that a 
theoretical understanding of the empirical correlation 
between SN Ia light curve shape and maximum luminosity may 
be emerging.  If the high value $q_0 \sim 0.5$ in the 
preliminary report \cite{Perlmutter96} is right, 
$\Omega_\Lambda$ is probably small and $\Omega_0\sim 1$. 

The era of structure formation is another important 
discriminant between these alternatives, low $\Omega$ 
favoring earlier structure formation, and $\Omega=1$ 
favoring later formation with many clusters and larger-scale 
structures still forming today.  A particularly critical 
test for models like CHDM is the evolution as a function of 
redshift of $\Omega_{gas}$ in damped Ly$\alpha$ systems. 

Reliable data on all of these issues is becoming available
so rapidly today that there is reason to hope that a clear
decision between these alternatives will be possible
within the next few years.

What if the data ends up supporting what appear to be
contradictory possibilities, e.g. large $\Omega_0$ {\it and}
large $H_0$?  Exotic initial conditions (e.g. ``designer''
primordial fluctuation spectra) or exotic dark matter
particles beyond the simple ``cold'' vs. ``hot''
alternatives (e.g. decaying intermediate mass neutrinos)
could increase the space of possible inflationary theories
somewhat.  But it may ultimately be necessary to go outside
the framework of inflationary cosmological models and
consider models with large scale spatial curvature, with a
fairly large $\Lambda$ as well as large $\Omega_0$.  This
seems particularly unattractive, since in addition to
implying that the universe is now entering a final
inflationary period, it means that inflation did not happen
at the beginning of the universe, when it would solve the 
flatness, horizon, monopole, and structure generation 
problems.  Therefore, along with most cosmologists, I am 
rooting for the success of inflation-inspired cosmologies, 
with $\Omega_0 + \Omega_\Lambda = 1$.  With the new upper 
limits on $\Lambda$ from gravitational lensing of quasars, 
number counts of elliptical galaxies, and high-redshift Type 
Ia supernovae, this means that the cosmological constant is 
probably too small to lengthen the age of the universe 
significantly.  So I am hoping that when the dust finally 
settles, $H_0$ and $t_0$ will both turn out to be low enough 
to be consistent with General Relativistic cosmology. But of 
course the universe is under no obligation to live up to our 
expectations.

\bigskip
\noindent ACKNOWLEDGEMENTS. I have benefited from conversations or
correspondence with S.  Bonometto, S.  Borgani, E. Branchini,
D. Caldwell, L. Da Costa, M. Davis, A. Dekel, S. Faber, G. Fuller, 
K. Gorski, K. Griest, J. Holtzman, A.  Klypin, C. Kochanek, K. Lanzetta,
P. Lilje, R. Mushotzky, R.  Nolthenius, J. Ostriker, P.J.E. Peebles,
S. Perlmutter, D.  Richstone, A. Sandage, R. Schild, J. Silk,
L. Storrie-Lombardi, R.B. Tully, E.L. Turner, M. Turner, A. Wolfe,
E. Wright, D.  York, and D. Zaritsky, and fruitful interactions with
UCSC graduate students R. Dav\'e, M. Gross, and R. Somerville.  This
research was supported by NASA, NSF, and UC research grants at UCSC.

\def\MNRAS{Mon.\ Not.\ R.\ Astron.\ Soc.}
\def\ApJ{Astrophys.\ J.}
\def\AJ{Astron.\ J.}
\def\AA{Astron.\ Astroph.}


\begin{thebibliography}{99}

\bibitem{Snowmass95}
This is a considerably revised version of my article in {\it 
Particle and Nuclear Astrophysics and Cosmology in the Next 
Millenium}, eds. E. Kolb and R. Peccei (Singapore: World 
Scientific, 1995), pp. 85-98.  A popular version is J. Roth 
and J.R. Primack, {\sl Sky and Telescope}, 91 (1), 20 
(January 1996). 

\bibitem{Freedman94}
W.L. Freedman et al., Nature 371, 757 (1994).

\bibitem{BFPR84}
G.R. Blumenthal, S.M. Faber, J.R. Primack, and M.J. Rees, Nature
311, 517 (1984); Erratum: 313, 72 (1985).  M. Davis, G. Efstathiou,
C.S. Frenk, and S.D.M. White, \ApJ\  292, 371 (1985).

\bibitem{Peeb84}
P.J.E. Peebles, \ApJ\ 284, 439 (1984).

\bibitem{CenOstGn}
L.A. Kofman, N.Y. Gnedin, and N.A. Bahcall, Astrophys. J.
413, 1 (1993);
R. Cen, N.Y. Gnedin, and J.P. Ostriker, Astrophys. J. 417,
387 (1993); R. Cen and J.P. Ostriker, Astrophys. J. 429, 4
(1994).

\bibitem{CHDM84}
S.A. Bonometto and R. Valdarnini, Phys. Lett. 103A, 369
(1984); L.Z. Fang, S.X. Li, S.P. Xiang, Astron. Astrophys.
140, 77 (1984); A. Dekel and S.J. Aarseth, Astroph. J.
283, 1 (1984); Q. Shafi and F.W. Stecker, Phys. Rev. Lett.
53, 1292 (1984).

\bibitem{DSS92}
M. Davis, F. Summers, and D. Schlegel, Nature 359, 393
(1992).

\bibitem{KHPR}
 A. Klypin, J. Holtzman, J.R. Primack, and E. Reg\H{o}s,
 \ApJ\ 416, 1 (1993).

\bibitem{Copi95}
C.J. Copi, D.N. Schramm, M.S. Turner, Science 267, 192
(1995).  Cf. also R.A. Malaney and G.J. Mathews, Phys.\ Rep.
229, 145 (1993); L.M. Krauss and P.J. Kernan, Phys. Lett.
B, 347, 347 (1995); N. Hata et al., Phys. Rev. Lett. 75,
3977 (1995); C.J. Copi, D.N. Schramm, and M.S. Turner, Phys.
Rev. Lett. 75, 3981 (1995); B.D. Fields and K. Olive, Phys.
Lett. B, 368, 103 (1996).

\bibitem{Dress87}
A. Dressler et al., \ApJ\  313, L37 (1987);  D. Lynden-Bell et al.,
\ApJ\ 326, 19 (1988).

\bibitem{Olivier93}
See e.g. S. Olivier, J.R. Primack, G.R. Blumenthal, and A.
Dekel, Astroph. J. 408, 17 (1993); A. Klypin and G. Rhee,
Astroph. J. 428, 399 (1994); and references therein.

\bibitem{White93}
S.D.M. White, G. Efstathiou, and C.S. Frenk, \MNRAS\  262, 1023
(1993).

\bibitem{Holtz89}
J. Holtzman, \ApJ\ Supp. 71, 1 (1989).

\bibitem{Wright92}
E.L. Wright, \ApJ\  396, L13 (1992).

\bibitem{HP93}
J.A. Holtzman and J.R. Primack, \ApJ\ 405, 428 (1993). These
results were presented (in J.R. Primack and J.A. Holtzman,
in {\it Gamma Ray --- Neutrino Cosmology and Planck Scale
Physics: Proc. 2nd UCLA Conf.}, February 1992, D. Cline, ed.
(World Scientific, 1993), pp. 28-44) before the anouncement
of the COBE discovery.  Cf. also T. van Dalen and R.K.
Schaefer, \ApJ\ 398, 33 (1992); R.K. Schaefer and Q. Shafi,
Phys. Rev. D 47, 1333 (1993).

\bibitem{BolteHogan}
M. Bolte and C.J. Hogan, Nature, 376, 399 (1995).

\bibitem{NewGC}
Note however that while R. Jimenez et al., astro-ph/9602132,
\MNRAS\ in press (1996) gives a best estimate for  the ages
of the oldest GCs of $13.5\pm2$ Gyr, they also give a lower
limit of 9.7 Gyr. A new analysis by M. Salaris, S.
Degl'Innocenti, A. Weiss, astro-ph/9603092 (1996),
also gives 13 Gyr as the best estimate of the ages of the
three old GCs that they studied (M15, M68, M92), and says
that a lower age is possible.

\bibitem{Chaboyer94}
B. Chaboyer, \ApJ\ 444, L9 (1995)
includes a table showing the effects on the GC ages of more
than 20 possible changes in input physics and abundances;
a Monte Carlo study (B. Chaboyer et al. 1995,
astro-ph/9509115) of variations in these parameters gives a
95\% CL lower limit on the ages of the oldest GCs of 12 Gyr.

\bibitem{Shi94}
X. Shi, \ApJ\ 446, 637 (1995).

\bibitem{Cosmochron}
R.A. Malaney, G.J. Mathews, and D.S.P. Dearborn,
\ApJ\ 345, 169 (1989); G.J. Mathews and D.N. Schramm,
\ApJ\ 404, 468 (1993).

\bibitem{Winget}
D.E. Winget  et al., Astrophys. J.  315, L77
(1987).

\bibitem{VdB}
 D.A. VandenBerg  et al., Astron. J.  100, 445
 (1990);  ibid., 102, 1043 (1991).

\bibitem{Yuan}
J.W. Yuan, Astron.\ Astrophys. 224, 108 (1989); 261, 105 (1992).  V.
Weidemann, Ann. Rev. Astron. Astrophys. 28, 103 (1990).
M.A. Wood, \ApJ\ 386, 539 (1992).  M. Hernanz  et al., \ApJ\ 434, 652 (1994).

\bibitem{Wood96}
T.D. Oswalt, J.A. Smith, and M.A. Wood, Nature, submitted.

\bibitem{SandageRev}
A. Sandage, in {\it Practical Cosmology: Inventing the
Past}, 23rd Sass Fee lectures, ed. B. Binggeli and R. Buser
(Berlin: Springer, 1995). A. Sandage and G.A. Tammann,
in {\it Advances in astrofundamental physics, Proc. Int. School of
Physics ``D. Chalonge''}, eds. N. Sanchez and A. Zichichi (World
Scientific, 1995).

\bibitem{SandageTam}
G.A. Tammann et al., astro-ph/9603076, to be pub.
in {\it Science with the HST - II} (1996).

\bibitem{Huchra92}
 J.P. Huchra, Science, 256, 321 (1992).
 S. Van den Bergh, Science 258, 421 (1992).

\bibitem{Jacoby92}
 G.H. Jacoby et al., Publ. Astron. Soc.  Pac.,
 104, 599 (1992).  R. Ciradullo, G.H. Jacoby, and J.L.
 Tonry, \ApJ\ 419, 479 (1993).

\bibitem{Fukugita93}
 M. Fukugita, C.J. Hogan, and P.J.E. Peebles, Nature
 366, 309 (1993).

\bibitem{KennFM}
R.C. Kennicutt, W.L. Freedman, and J.R. Mould, Astron. J.
110, 1476 (1995).

\bibitem{RR85}
 M. Rowan-Robinson, {\it The Cosmic Distance Ladder} (1985).

\bibitem{Pierce94}
 M.J. Pierce et al., Nature 371, 385 (1994).

\bibitem{vdBergh96}
S. van den Bergh, astro-ph/9509117 (1995); D.
Branch et al., astro-ph/9604006 (1996).  Cf. B.
Schaefer, \ApJ\ 459, 438 (1996).

\bibitem{Press94}
 M.M. Phillips, \ApJ\ 413, L105 (1993); M. Hamuy et al., Astron. J.
 109, 1 (1995).
 A.G. Riess, W.H. Press, and R.P. Kirshner, \ApJ\ 438, L17
 (1995); they get $h=0.67\pm0.07$.

\bibitem{SvsPhil}
G.A. Tammann and S. Sandage, \ApJ\ 452, 16 (1995).

\bibitem{Tonry}
 J. Tonry, \ApJ\ 373, L1 (1991).

\bibitem{Branch}
 W.D. Arnet, D. Branch, and J.C. Wheeler, Nature 314, 337 (1985);
D. Branch, Astroph. J. 392, 35 (1992).  For more recent results see
A. Fisher et al., \ApJ\ 447, L73; S. van den Bergh, astro-ph/9509117,
\ApJ\ in press.

\bibitem{Leibundgut}
 B. Leibundgut and  P.A. Pinto, Astroph. J. 401, 49 (1992).  But
 cf. T.E. Vaughan et al., \ApJ\ 439, 558 (1995).

\bibitem{Nugent}
P. Nugent, D. Branch, et al., Phys. Rev. Lett. 75, 394;
Erratum 75, 1874 (1995).  Reviews: D. Branch and A.M.
Khokhlov, Phys. Rep. 256, 53 (1995); D. Branch et al.,
astro-ph/9601006, to appear in {\it Proc. NATO
Advanced Studies Institute on Thermonuclear Supernovae},
Aiguablava, Spain, June, 1995, eds. R. Canal, P.
Ruiz-Lapuente, and J. Isern (1996).

\bibitem{SKE92}
 B.P. Schmidt, R.P. Kirschner, and R.G. Eastman, \ApJ\ 395,
 366 (1992).

\bibitem{Schmidt94}
 B.P. Schmidt et al., \ApJ\ 432, 42 (1994).  See also Robert
Kirschner's Varenna lectures.

\bibitem{SNIIenv}
M. Best and R. Wehrse, Astron. Astroph. 284, 507 (1994).  P.
Ruiz-Lapuente et al., \ApJ\ 439, 60 (1995).

\bibitem{Birk91}
 M. Birkinshaw, J.P. Hughes, and K.A. Arnoud, \ApJ\ 379, 466
 (1991).

\bibitem{Birk94}
 M. Birkinshaw and J.P. Hughes, \ApJ\ 420, 33 (1994).

\bibitem{ASCA}
 K. Yamashita, in {\it New Horizon of X-ray Astronomy ---
 First Reslts from ASCA}, eds. F. Makino and T. Ohashi
 (Universal Academy Press, Tokyo, 1994), p. 279.

\bibitem{Lange}
 T.M. Wilbanks et al., \ApJ\ 427, L75 (1994), and parallel
 session talk at Snowmass 94.

\bibitem{Raphaeli}
 Y. Rephaeli, Ann. Rev. Astron. Astrophys 33, 541 (1995).

\bibitem{Pelt94}
 J. Pelt et al., Astron.\ Astroph. 286, 775 (1994).

\bibitem{Press92}
 W.H. Press, G.B. Rybicki, and J.N. Hewitt, \ApJ\ 385, 416
 (1992).

\bibitem{ETurner95}
T. Kundic et al., \ApJ\ 455, L5 (1995).

\bibitem{Rhee91}
 G.F.R.N. Rhee, Nature 350, 211 (1991).

\bibitem{Roberts91}
 D.H. Roberts et al., Nature 352, 43 (1991).

\bibitem{Danle}
 H. Danle, S.J. Maddox, and P.B. Lilje, \ApJ\ 435, L79 (1994).
Also further analyses in prep.

\bibitem{Narayan}
N.A. Grogan and R. Narayan, \ApJ\ in press (1996).


\bibitem{Mould95}
J. Mould et al., Astrophys. J. 449, 413 (1995).  Cf. also N.R. Tanvir
et al., astro-ph/9509160, which uses new HST Cepheids in M96 to
calibrate early-type galaxies and deduce a distance to the Coma
cluster, leading to $H_0=69\pm8$.

\bibitem{lowH}
 E.L. Turner, R. Cen, and J.P. Ostriker, Astron. J. 103,
1427 (1992); X.-P. Wu et al., preprint (1995).

\bibitem{Suto94}
 Y. Suto, T. Suginohara, and Y. Inagaki, Prog.\ Theor.\ Phys.\
 93, 839 (1995).

\bibitem{KimLBL}
A. Kim et al., astro-ph/9602123, to appear
in {\it Thermonuclear Supernovae (NATO ASI)}, eds. R. Canal,
P. Ruiz-LaPuente, and J. Isern (1996).

\bibitem{RatraP}
 M. Sasaki, et al. Phys. Lett. B 317, 510 (1993).
 B. Ratra and P.J.E. Peebles, \ApJ\ 432, L5 (1994); Phys. Rev. D 52,
 1837 (1995).   M.  Kamionkowski et al., \ApJ\ 434, L1 (1994).
 M. Sasaki, T. Tanaka, and K. Yamamoto, Phys. Rev. D 51, 2979 (1995).
 M. Bucher, A.S.  Goldhaber, and N. Turok, Phys. Rev. D 52, 3314 (1995).

\bibitem{LLPR}
 O. Lahav, P. Lilje, J.R. Primack, and M.J. Rees,
 \MNRAS\  251, 128 (1991).

\bibitem{LambdaARAA}
 S.M. Carroll, W.H. Press, and E.L. Turner, Ann. Rev.
 Astron. Astrophys, 30, 499 (1992).

\bibitem{MaozRix}
 D. Maoz and H.W. Rix, \ApJ\ 416, 425 (1993).

\bibitem{Kochanek94}
 C. Kochanek, \ApJ\ 419, 12 (1993).

\bibitem{Kochanek95}
 C.S. Kochanek, astro-ph/9510077, \ApJ\ in press (1996).

\bibitem{SCarlberg}
C. Steidel, M. Dickinson, and S.E. Persson, \ApJ\ 437, L75 (1994);
S. Lilly et al., \ApJ\ 455, 108 (1995);
D. Schade et al., astro-ph/9604032 (1996).

\bibitem{BunnS94}
 E. Bunn and N. Sugiyama, Astrophys. J. 446, 49 (1995).
However, this limit may be stronger than is justified by a
more complete analysis of the COBE data (E. Wright and K.
Gorski, private communications).

\bibitem{Driver96}
S. Driver et al., astro-ph/9511141, \ApJ\ in press
(April 1996).

\bibitem{KPH96}
 A. Klypin, J.R. Primack, and J. Holtzman, astro-ph/9510042,
\ApJ\ in press (July 1996).

\bibitem{Perlmutter95}
 G. Goldhaber et al., Nucl. Phys. B, S38, 435 (1995).
 A. Goobar and S. Perlmutter, Astrophys. J. 450, 14 (1995).
 S. Perlmutter et al., Astrophys. J. 440, L41 (1995).

\bibitem{Perlmutter96}
 S. Perlmutter, astro-ph/9602122, to appear in {\it
Thermonuclear Supernovae (NATO ASI)}, eds. R. Canal, P.
Ruiz-Lapuente, and J. Isern (1996).

\bibitem{DekelARAA}
 A. Dekel, Ann. Rev. Astron. Astroph. 32, 371 (1994).
 M.A. Strauss and J.A. Willick, Phys. Rep. 261, 271 (1995).
 The latest POTENT results on the power spectrum of mass
 fluctuations are in S. Zaroubi et al., astro-ph/9603068 (1996)
 and T. Kolatt and A. Dekel, astro-ph/9512132 (1995).

\bibitem{DekRees}
 A. Dekel and M.J. Rees, \ApJ\ 422, L1 (1994).

\bibitem{NusDek}
 A. Nusser and A. Dekel, \ApJ\ 405, 437 (1993).  Similar
constraints have been derived using a perturbative technique
by F. Bernardeau, R. Juskiewicz, A. Dekel, and F.R. Bouchet,
\MNRAS\ 274, 20 (1995).

\bibitem{Peebles}
P.J.E. Peebles, {\it Principles of Physical Cosmology}
(Princeton Univ. Press, 1992), esp. \S20.

\bibitem{DavisP}
M. Davis and P.J.E. Peebles, \ApJ\ 267, 465 (1983).

\bibitem{Somerville}
R. Somerville, M. Davis, and J.R. Primack,
astro-ph/9604041, \ApJ\ submitted (1996).

\bibitem{Moetal}
H.J. Mo, Y.P. Jing, and G. B\"orner, \MNRAS\ 264, 825 (1993).
W. Zurek {\it et al.}, \ApJ\ 431, 559 (1994).

\bibitem{Fisher94}
K.B. Fisher, et al., \MNRAS\  267, 927 (1994).

\bibitem{Marzke}
R.O. Marzke, M.J. Geller, L.N. da Costa, and J.P. Huchra,
Astron. J. 110, 477 (1995).

\bibitem{Somerville2}
R. Somerville, J.R. Primack, and R. Nolthenius,
astro-ph/9604051, \ApJ\ submitted (1996).

\bibitem{KOS79}
R. Kirschner, A. Oemler, and P. Schechter, AJ 84, 951
(1979).

\bibitem{HuchraG82}
J.P. Huchra and M.J. Geller, \ApJ\ 257, 423 (1982).

\bibitem{Ramella89}
M. Ramella, M.J. Geller, and J.P. Huchra, \ApJ\ 344, 57
(1989).

\bibitem{NKP95}
R. Nolthenius, A. Klypin, and J.R. Primack,
astro-ph/9410095, \ApJ, in press (1996).

\bibitem{CarlCouch}
R.G. Carlberg and H.M.P. Couchman, \ApJ\ 340, 47 (1989).

\bibitem{NKP94}
R. Nolthenius, A. Klypin, and J.R. Primack, \ApJ\ 422, L45
(1994).

\bibitem{PeebAction}
P.J.E. Peebles, \ApJ\ 344, 53 (1989); 362, 1 (1990); 429, 43
(1994).

\bibitem{BranCarl}
E. Branchini and R.G. Carlberg, \ApJ\ 434, 37 (1994).

\bibitem{DunnLaf}
A.M. Dunn and R. Laflamme, \ApJ\ 443, L1 (1995).

\bibitem{TullyPS}
E.J. Shaya, P.J.E. Peebles, and R.B. Tully, \ApJ\ 454, 15 (1995).

\bibitem{Zaritsky}
D. Zaritsky et al., \ApJ\ 405, 464 (1993); D. Zaritsky and
S.D.M. White, \ApJ\ 435, 599 (1994).

\bibitem{ZaritskyPC}
D. Zaritsky, private communication (1995).

\bibitem{Dubinski96}
J. Dubinski, J.C. Mihos, and L. Hernquist,
astro-ph/9509010, \ApJ\ in press (1996).

\bibitem{Tyson84}
J.A. Tyson et al., \ApJ\ 281, L59 (1984); cf. I. Kovner and
M. Milgrom, \ApJ\ 321, L113 (1987).

\bibitem{MouldBland}
J. Mould et al., \MNRAS\ 271, 31 (1994).

\bibitem{Villum95}
J. Villumsen, astro-ph/9507007 (1995).

\bibitem{LitTre87}
B. Little and S. Tremaine, \ApJ\ 320, 493 (1987).

\bibitem{KochanMWy}
C.S. Kochanek, astro-ph/9505068, submitted to \ApJ\ (1995).

\bibitem{Songaila}
A. Songaila et al., Nature 368, 599 (1994); R.F. Carswell et
al. \MNRAS\  268, L1 (1994).

\bibitem{Tytler}
D. Tytler and X. Fan, Bull.\ Am.\ Astron.\ Soc. 26, 1424 (1994);
D. Tytler, X. Fan, and S. Burles, astro-ph/9603069, submitted to
Nature (1996).  S. Burles and D. Tytler, astro-ph/9603070, submitted
to Science (1996).

\bibitem{Hogan95}
 M. Rugers and C.J. Hogan, \ApJ\ 459, L1 (1996).
 cf. C.J. Hogan, astro-ph/9512003 (1995).

\bibitem{Wampler}
E.J. Wampler et al., astro-ph/9512084, submitted to Astron.
Astroph. (1995).

\bibitem{Carswell96}
R.F. Carswell et al., \MNRAS\ 278, 506 (1996).

\bibitem{WhiteClus}
S.D.M. White and C.S. Frenk, \ApJ\ 379, 52 (1991).  S.D.M.
White et al. Nature 366, 429 (1993).  Cf. Lubin et al.,
\ApJ\ 460, 10 (1996); A.E. Evrard, C.A. Metzler, and
J.F. Navarro, FERMILAB-Pub-95/337-A (1995); S. Schindler, \AA\
305, 756 (1996).  Review: G. Steigman and J.E. Felten,
Space Science Reviews 74, 245 (1995).

\bibitem{Fabian}
D.A. White and A.C. Fabian, \MNRAS, 273, 72 (1995).  Cf. R. Mushotzky
et al., in {\it Dark Matter}, ed. S.S. Holt and C.L. Bennett, AIP
Conf. Proc. 336, 231 (1995); D.A. Buote and C.R. Canizares, \ApJ\ 457,
565 (1996); S. Bardelli et al., \AA\ 305, 435 (1996).  However, recent
ASCA data on several rich clusters imply lower baryon fractions,
less than 0.1 for $h=0.5$ (R. Mushotzky private communication, March
1996), which may call into question the ``fair sample'' hypothesis.

\bibitem{ClusBar}
K.F. Gunn and P.A.  Thomas, astro-ph/9510082 (1995) show that the
baryon fraction could be overestimated by a factor of 2 or more if the
cluster gas is a multiphase medium.

\bibitem{LoebMao}
A. Loeb and S. Mao, \ApJ\ 435, L109 (1994).  P. Biermann (private
communication) has suggested that such tangled fields might extend
throughout clusters; cf. J.E. Felten, in {\it Clusters, Lensing, and
the Future of the Universe}, eds. V. Trimble and A. Reisenegger,
ASP Conf. Series 88, 271 (1996).

\bibitem{Kronberg}
P.P. Kronberg, Reports on Progress in Physics 57, 325 (1994).

\bibitem{BBlanchard}
C. Balland \& A. Blanchard, astro-ph/9510130 (1995).

\bibitem{BLubin}
N.A. Bahcall and L.M. Lubin, \ApJ\ 426, 513 (1994).  M. Bartelmann
and R. Narayan, in {\it Dark Matter}, College Park, MD, October 1994
(AIP Conference Proceedings 336, 1995) 307.

\bibitem{Kaiser}
G. Falhman et al. \ApJ\ 437, 56 (1994), reviewed in N. Kaiser et al.,
astro-ph/9407004 (1994).  There are several similar examples.  However,
R.G. Carlberg, H.K.C. Yee, and E. Ellingson,  astro-ph/9512087 (1995)
show that the velocities of galaxies around 16 galaxy
clusters are probably not consistent with such large dark matter halos.
This puzzling discrepancy needs further study.

\bibitem{BHNPK}
This is shown visually (with accompanying video; high-resolution
frames and low-resolution mpeg excerpts
from the video can be viewed or downloaded from our WWW page
http: //phy\-sics.ucsc.edu/groups/cos\-mo\-lo\-gy.html)
in my group's CHDM simulations in D. Brodbeck
et al., \ApJ\ in press (1996); cf. also A. Klypin,
R. Nolthenius, and J.R. Primack, astro-ph/9502062, \ApJ\ in press (1996).
Observed cluster temperatures and even X-ray luminosities appear
to be well fit,
according to the first CHDM hydrodynamic cluster simulation, G. Bryan
et al., \ApJ\ 437, L5 (1994).

\bibitem{KofmanKP}
 L. Kofman, A. Klypin, D. Pogosyan, and J.P. Henry, astro-ph/9509145
 (1995).

\bibitem{Richstone92}
D. Richstone, A. Loeb, and E.L. Turner, \ApJ\ 393, 477 (1992).

\bibitem{Burns}
J.O. Burns et al., \ApJ\ 427, L87 (1994).

\bibitem{Mohr95}
J.J. Mohr, A.E. Evrard, D.G. Fabricant, and M.J. Geller, \ApJ\
447, 8 (1995).

\bibitem{Jing95}
Y.P. Jing, H.J. Mo, G. B\"orner, and L.Z. Fang, in {\it International
Workshop on Large Scale Structure in the Universe}, Potsdam, 1994
(World Scientific, 1995);  \MNRAS\ 276, 417 (1995).

\bibitem{Tsai}
E.g., D.A. Buote and J.C. Tsai, \ApJ, 458, 27 (1996), and J.C. Tsai
and D.A. Buote, astro-ph/9510057, \MNRAS\ in press (1996).

\bibitem{Gioia}
J.P. Henry et al., \ApJ\ 386, 408 (1992). F.J. Castander et al.,
Nature 377, 39 (1995).  H. Ebeling et al., to appear in {\sl
Roentgenstrahlung from the Universe}, Wurzberg, Sept. 1995.

\bibitem{Steidel94}
C.C. Steidel, M. Dickinson, and S.E. Persson, \ApJ\ 437,
L75 (1994).

\bibitem{Kauffmann94}
G. Kauffmann, \MNRAS\ 274, 161 (1995).

\bibitem{Klypin95}
A. Klypin et al., in prep. (1996).

\bibitem{Haehnelt}
M.G. Haehnelt, \MNRAS\  265, 727 (1993).

\bibitem{Katz94}
N. Katz, T. Quinn, E. Bertschinger, and J.M. Gelb, \MNRAS\
270, L71. Cf. D.J. Eisenstein and A. Loeb, \ApJ\ 443, 11 (1995).

\bibitem{Steidel96}
C.C. Steidel et al., astro-ph/9602024 (1996), and M. Giavalisco,
C.C. Steidel, and P.D. Macchetto, astro-ph/9603062 (1996).

\bibitem{Wolfe}
A.M. Wolfe, in {\it Relativistic Astrophysics and Particle
Cosmology}, eds. C.W. Ackerlof, M.A. Srednicki (New York,
New York Academy of Science, 1993), p. 281.

\bibitem{LWT}
K.M. Lanzetta,  Publ/\  Astron.\  Soc.\ Pac.,
105, 1063, (1993);
K.M. Lanzetta, A.M. Wolfe and D.A. Turnshek, \ApJ\ 440, 435 (1995).

\bibitem{MoME}
H.J. Mo and J. Miralda-Escude, \ApJ\ 430, L25 (1994);
G. Kauffmann and S. Charlot, \ApJ\ 430, L97 (1994);
C.-P. Ma and E. Bertschinger, \ApJ\ 434, L5 (1994).

\bibitem{KBHP}
A. Klypin, S. Borgani, J. Holtzman, and J.R. Primack, \ApJ\ 444, 1
(1995).

\bibitem{Storrie}
L.J. Storrie-Lombardie, R.G. McMahon, M.J. Irwin, and C.  Hazard,
\ApJ\ 427, L13-16 (1994); ----, {\it Proc. ESO Workshop on QSO
Absorption Lines}, ed. G. Meylan (Springer, 1995).  Newer data suggest
that $\Omega_{gas}\approx 3\times 10^{-3}$ for $z\approx 2-3.5$, and
either declining or at most remaining constant for higher $z\sim
4-4.5$ (private communications from Lisa Storrie-Lombardi and Art
Wolfe, December 1995).

\bibitem{LLu}
L.M. Lu et al., \ApJ\ 457, L1 (1996).

\bibitem{Haehnelt96}
M. Haehnelt, M. Steinmetz, and M. Rauch, astro-ph/9512118 (1996).

\bibitem{Whipple96}
J. Quinn et al., \ApJ\ 456, L83 (1996); M.S. Schubnell et al.,
astro-ph/9602068, \ApJ\ in press (1996).

\bibitem{MacMinnP}
D. MacMinn and J.R. Primack,
in {\it TeV Gamma Ray Astrophysics}, ed. Heinz V\"olk and
F. Aharonian, Space Science Reviews, 75, 413 (1996).
Cf. Ground Based Gamma-Ray Astronomy,
in {\it Particle and
Nuclear Astrophysics and Cosmology in the Next Millenium},
eds. E. Kolb and R. Peccei (Singapore: World Scientific,
1995), p. 295.

\bibitem{CaldwellStock}
D.O. Caldwell, in {\it Trends in Astroparticle Physics},
Stockholm, Sweden 22-25 September 1994, eds. L. Bergstrom,
P. Carlson, P.O. Hulth and N. Snellman, Nucl. Phys. B, Proc.
Suppl., 43, 126 (1995).  C. Athanassopoulos et al.,
Phys. Rev. Lett. 75, 2650 (1995).
In the latest preprint from LSND, available at
http://nu1.lampf.lanl.gov/~lsnd/, the 1995 data strengthens
the statistics implying oscillations, but perhaps weakens 
the case for $\delta m^2=6$ eV$^2$.

\bibitem{Fukuda}
Y. Fukuda, Phys. Lett. B 335, 237 (1994).

\bibitem{PHKC}
J.R. Primack, J. Holtzman, A. Klypin, and D.O. Caldwell,
in {\it Trends in Astroparticle Physics},
Stockholm, Sweden 22-25 September 1994, eds. L. Bergstrom,
P. Carlson, P.O. Hulth and N. Snellman, Nucl. Phys. B, Proc.
Suppl., 43, 133 (1995), and
Phys. Rev. Lett. 74, 2160 (1995); J.R. Primack, J. Holtzman,
and A. Klypin, Moriond 1995, in press.
Cf. D.Yu. Pogosyan and A.A. Starobinsky, in {\it International
Workshop on Large Scale Structure in the Universe}, Potsdam, 1994
(World Scientific, 1995) and \ApJ\ 447, 465 (1995);
K.S. Babu, R.K. Schaefer, and Q. Shafi, Phys. Rev. D53, 606
(1996); A.R. Liddle et al., astro-ph/9511057 (1995).

\bibitem{FPQ}
However, an inverted neutrino
mass hierarchy with $\nu_e$ the most massive appears to be
required if all the current experimental evidence for
neutrino mass (solar and atmospheric neutrino deficits, and
LSND) are valid, and r-process nucleosynthesis (responsible
for production of the heavy elements) takes place in the hot
$\nu$ bubble a few hundred km above the neutron star in Type
II supernovae; see G.M. Fuller, J.R. Primack, and Y. Qian,
Phys. Rev. D 52, 1288 (1995).

\bibitem{DGS95}
See e.g. S. Dodelson, E. Gates, and A. Stebbins, astro-ph/9509147
(1995); G. Jungman et al., astro-ph/9512139 (1995); M. Tegmark,
astro-ph/9601077, \ApJ\ Lett. in press (1996).

\end{thebibliography}
\end{document}